# Methods for estimating the size of Google Scholar

Enrique Orduna-Malea[1]*, Juan M Ayllón[2], Alberto Martín-Martín[2] and Emilio Delgado López-Cózar[2]

[1] *EC3 Research Group, Polytechnic University of Valencia. Camino de Vera s/n, Valencia 46022, Spain*
[2] *EC3 Research Group, Universidad de Granada, 18071 Granada, Spain*

* enorma@upv.es

**Abstract** The emergence of academic search engines (mainly Google Scholar and Microsoft Academic Search) that aspire to index the entirety of current academic knowledge has revived and increased interest in the size of the academic web. The main objective of this paper is to propose various methods to estimate the current size (number of indexed documents) of Google Scholar (May 2014) and to determine its validity, precision and reliability. To do this, we present, apply and discuss three empirical methods: an external estimate based on empirical studies of Google Scholar coverage, and two internal estimate methods based on direct, empty and absurd queries, respectively. The results, despite providing disparate values, place the estimated size of Google Scholar at around 160-165 million documents. However, all the methods show considerable limitations and uncertainties due to inconsistencies in the Google Scholar search functionalities.

## 1. Introduction

Being able to access the entire body of human academic knowledge remains a long-standing dream since this is an engine for the advancement of science (Berman 2012). The emergence of global academic search engines (Ortega 2014) has revived this interest and made us wonder to what extent this wealth of knowledge is indexed, searchable and freely accessible online. Gauging its extent necessarily involves ascertaining the coverage of these sources, even though we are aware they constitute only the tip of an iceberg (Khabsa and Giles 2014).

In traditional bibliographic databases (WoS, Scopus), finding out the size (measured by the number of records at a given time) is a fairly trivial matter (a direct query will inform us that Web of Science Core Collection holds 56.9 million records and Scopus 54.5, as of May 2014), because the entire online universe is catalogued and under control (always accounting for a low error rate due to a lack of absolute normalisation in the catalogue). Moreover, the evolution of these databases is cumulative, i.e., the number of records always grows and never decreases, except for the occasional elimination of records due to technical or legal issues.

However, in the case of academic search engines, these assertions do not always apply (Ortega, Aguillo and Prieto 2006; Payne and Thelwall 2008; Orduna-Malea, Serrano-Cobos and Lloret-Romero 2009), making both calculating their size and tracking the evolution of their data a hard task, due essentially to the following issues.

On the one hand, the universe of catalogued records on a pure academic search engine depends directly on World Wide Web public coverage, and is therefore affected by its high dynamism (Brewington and Cybenko 2000; Adar, Teevan and Dumais 2009) and growth rate (Barabasi and Albert 1999; Adamic and Huberman 2001; Levene, Fenner, Loizou and Wheeldon 2002). Contents are continually added, changed and/or deleted worldwide (Koehler 1999; 2002; 2004), and these issues give rise to inherent technical difficulties in cataloguing and updating such a vast, diverse and complex universe.

On the other hand, there is a heavy dependence not only on the search engine functionalities but also on the information and web policies followed by those responsible for the databases. In this sense, there are some features sorely missing in Google Scholar, such as the availability of an Application Programming Interface (API), the existence of advanced filtering options, grouping of the different versions of the same academic contribution, classification by discipline, or the availability of bibliometric information, among others (Ortega 2014). Moreover, the inclusion of documents in Google Scholar partially depends on agreements with the publishers who hold the intellectual property rights for the documents.

However, despite the methodological problems of measuring these databases, the approximate calculation of their size and evolution sheds light on the patterns of creation, dissemination and use of scientific literature through the Web. For example, Microsoft Academic Search is essential for conducting retrospective studies because it gathers scientific literature since 1700, whereas Web of Science Core Collection (WoSCC) does not provide data until 1898, and Scopus until 1823 (although its coverage is highly irregular until the 1990s).

In the case of Google Scholar, it not only has a wider retrospective coverage than WoSCC or Scopus, but it also covers a wider range of languages and publication types (Meho and Yang 2007), and its growth rate is higher (Harzing 2014; Orduna-Malea and Delgado López-Cózar 2014). Moreover, it retrieves a higher number of citations, the intensity of which varies depending on the area of knowledge (Kousha and Thelwall 2008).

Furthermore, the use of Google Scholar by students and researchers as the main source for searching and using scientific literature is constantly growing (JISC 2012; Van Noorden 2014). A perceived usefulness and ease of use are the key reasons that contribute to the student's intention to use Google Scholar (Cothran, 2011).

Nonetheless, the use of Google Scholar for bibliometric purposes provides a new number of shortcomings (Jacso 2008; Aguillo 2011), such as lack of accuracy, error in cataloguing, attributing erroneous citations or including not strictly academic material, which should be emphasised as well. In addition to this, we must consider that the coverage of Google Scholar is also limited to what its parsers can crawl on the public Web (or under agreements with publishers). In this sense, WoS is much superior for old contributions which exist only in paper form, or which are not scanned and (properly) digitised.

Khabsa and Giles (2014) recently estimated the number of circulating documents written in English on the academic web at 114 million (of which GS has around 99.8 million) They employed an innovative procedure based on the Lincoln-Petersen (capture-recapture) method, obtaining incoming citations to a sample of articles written in English included both in Google Scholar and Microsoft Academic Search.

The advantage of this method is that it relies not on Google Scholar search functionalities but on a mixture of bibliometric and webometric techniques, thus avoiding Google Scholar search issues. However, the procedure depends on the precision of the Lincoln-Petersen method (employed through an analogy), and has been tested only on English and article-type documents.

These results lead us to formulate the following research question: is it possible to calculate the global size (the number of indexed documents) of Google Scholar considering all languages and document types?

Therefore, the main objective of this article is to propose and apply various methods to estimate the size of Google Scholar, pointing up their strengths and weaknesses.

## 2. Research background

Calculating the size of the Internet, in general, and the Web, in particular (Lawrence and Giles 1998; 1999; Albert, Jeong and Barabasi 1999; Dobra and Fienberg 2004) has generated a debate in the scientific arena over the last two decades. This debate has focused on various aspects, including the following.

First, from a socioeconomic perspective: how the composition and evolution of content affect the way it is consumed in different countries according to varying social, economic and political issues. For example, The Web Index[1], created by the World Wide Web Consortium in 2012, aims to measure the Web's growth, utility and impact on people and nations through a multidimensional ranking system, measuring universal access, relevant content, freedom and openness, and impact and empowerment dimensions.

Second, from an information perspective: the extent to which all the knowledge produced is actually indexed, searchable, retrievable and accessible from a catalogue, index or database. Both the processes of digitisation and the web policies related to the creation and dissemination of scientific- and academic-related material fall into this category.

Finally, from a methodological perspective: the number of external variables that should be taken into consideration to calculate content as accurately as possible. The opacity of multimedia files (storing text as images in PDF files), the difficulty of managing dynamic URLs, the influence of URL shorteners, the dynamism of the Web or the dependence on search engine functionalities are some of the methodological problems, and have recently been summarised from a webometric perspective by Wilkinson and Thelwall (2013).

The choice of the unit of analysis is another important issue. In the case of academic search engines (especially Google Scholar and Microsoft Academic Search), the unit of measurement is the document (journal article, conference, book, etc.). As regards Google Scholar, a "document" should be understood in principle as an academic document hosted in the "academic web". On the one hand, an academic document consists of an electronic file (essentially HTML and PDF) which contains journal and conference papers, theses and dissertations, academic books, pre-prints, abstracts, technical reports and other scholarly literature. This document must present an academic structure (title, author, abstract, and references) to be successfully indexed. On the other hand, the academic web consists of trusted web sites, supposed to host academic materials (such as universities, repositories, libraries, academic publishers, professional societies, academic databases, academic search engines, academic social networks, etc.). Additionally, any webmaster may request to be indexed on Google Scholar, i.e., institutions or services not directly oriented to research activities (such as personal web pages, private companies, blogs, etc.).

Additionally, Google Scholar includes additional items: a) citations (references to academic documents not indexed yet in the database), and b) other non-academic documents (patents and case laws), which are integrated in the database directly (this issue will be further developed in the Method section). All of them (academic documents, citations, and non-academic documents) are considered "Google Scholar documents" in this research, and conform the total size of Google Scholar.

Although there is also high dynamism on academic web resources (Ortega, Aguillo and Prieto 2006; Payne and Thelwall 2007; 2008a; 2008b), the processes of creation and dissemination of scientific material are governed by different processes, thus allowing greater control.

Nevertheless, studies focusing on academic search engines have been scarce, and have focused primarily on citations gathered from the analysis of certain units, such as institutions (Yang and Meho 2006), document types (Meho and Yang 2007; Winter, Zadpoor and Dodou 2014; Delgado López-Cózar and Cabezas-Clavijo 2013) or disciplines (Miri, Raoofi and Heidari 2012; Kousha and Thelwall 2008).

The total quantification of academic search engines has barely been studied. With regard to Microsoft Academic Search, we might note the work of Jacsó (2011) and Orduna-Malea, Martín-Martín, Ayllón and Delgado López-Cózar (2014). In addition, Ortega (2014) has provided quantitative data from a wide range of search engines, although some of these are already outdated or have disappeared entirely (like Scirus).

Regarding Google Scholar, the largest academic search engine today, it is worth mentioning the work of Aguillo (2012), who estimated its size at 83 million documents (as of 2010) through an analysis of TLDs (Top Level Domains). Subsequently, Khabsa and Giles (2014) put the size of Google Scholar at 99.3 million documents (as of January 2013), while Ortega (2014) puts the figure at 109.3 million (as of December 2013), of which 94.73 million correspond to scholarly documents, while the remaining records correspond to court opinions and case law, not included in previous studies.

While Aguillo (2012) and Ortega (2014) use internal methods (using the search functionalities of Google Scholar), Khabsa and Giles (2014) use an external method (an estimate from external parameters), and focus only on English documents. The coverage is thus different in each study, which hinders direct comparisons.

It therefore follows that there is a need to address the size of Google Scholar holistically (considering all document types included), employing various empirical methods, both internal (using the search tools Google Scholar provides) and external (using tools from outside Google Scholar).

## 3. Method

In order to estimate the size of Google Scholar (number of indexed documents) we propose and test three different procedures, one external (not directly using Google

Scholar functionalities) and two internal methods (using Google Scholar's search functionalities), explained in further detail below.

### 3.1. External method: estimates from empirical studies of Google Scholar coverage

The first method consists of making estimates from empirical studies that have previously used small samples and have compared GS with other databases. From these comparisons (paying special attention to differences in coverage), a correction factor may be obtained, and consequently a hypothetical projection proposed.

To this end, an extensive collection of empirical studies on how to calculate the sizes of academic databases was gathered (as of May 2014; available in the supplementary material, Appendix I). For each collected work, we provide the bibliographic database analysed (GS, WoS, MAS, Scopus, Pubmed, etc.), the sample size considered, and finally the unit of analysis.

These studies use different units of analysis (journals, articles, books, etc.) and measures (citation count, h-index, Thomson Reuters' Impact Factor, etc.). Nonetheless, for our research purposes, we decided only to apply the synthesis of the results to samples that are comparable to each other on the following levels:
- Studies examining the same databases as their data source.
- Studies working with documents or with unique citation documents.
- Studies that make comparisons between documents written in the same language (or do not make a distinction by language).

Hence, and in order for comparisons to be feasible, we categorised the data provided in Appendix I according to a) the unit of study: journals, books, etc. (Appendix II); b) the measured indicator: citations, documents, citations per document (Appendix III); and c) the language of the documents (Appendix IV).

Only those studies comparing GS and WoS were considered, since only two studies provided information about empirical comparisons between GS and Scopus, and the remaining databases were even less represented.

Next, for each case study we obtained a proportion factor (PF) between the two databases (GS and WoS) by dividing the number of documents retrieved on GS by the number of documents retrieved on WoS. Finally, the median of all the studies was calculated to obtain a rough, but indicative, correction factor (CF):

**Table 1. Correction factor calculation**

| EQUATION | OBSERVATIONS |
|---|---|
| $PF_i = \dfrac{x_i}{y_i}$ | $x_i$ = n° of documents retrieved on GS in empirical study number "i". $y_i$ = n° of documents retrieved on WoS in empirical study number "i". $PF_i$ = Proportion factor in empirical study number "i". |
| $PF_1, PF_2, PF_3 \cdots, PF_n$ | $PF_i$ are sorted from highest to lowest value; n= n° of empirical studies. |
| $CF = \dfrac{PF_{i=n+1}}{2}$ | $CF$ = Correction factor; if n = odd |
| $CF = \dfrac{PF_{i=\frac{n}{2}} + PF_{i=\frac{n}{2}+1}}{2}$ | $CF$ = Correction factor; if n= even |

This same procedure was also applied to the comparison of unique citing documents as unit of analysis (citing documents indexed in GS and non-indexed in WoS, and vice versa).

**3.2. Internal method**

The second method is based on interrogating the database itself by using the functionalities provided by Google Scholar, at least to the extent that this is possible. In this sense, this method relies on the quantification of "hits".

Theoretically, for each academic document hosted in the academic web, and successfully indexed in the Google Scholar database, a bibliographic record is created. When a query is submitted through the search box, the search engine results page (SERP) shows a number of documents that match the query. Each document is presented by means of its bibliographical record, and each record is considered a "hit". Thus, interrogating the database to obtain as many hits as possible may be a way to retrieve the number of bibliographic records created, being this figure an approximation of the number of documents indexed in Google Scholar (called size in this research).

At this point it should be noted that all documents indexed in Google Scholar are not strictly academic documents stored in the academic web. Google Scholar gives two different types of result (articles and case laws). Articles in turn comprise three types of result (some of which may be included or excluded in a query): ordinary records (academic documents with abstract indexed on Google Scholar, which may provide a link to the full text or to a paid gateway); citations (references to academic documents not indexed on Google Scholar); and patents (documents extracted from Google Patents).

Under this direct method, the size of Google Scholar is thus interpreted as the total number of hits (articles + case laws), including patents and citations.

**a) Direct query**

This can be done by two procedures: a) using the custom date range for the complete period of time (wide year range); and b) using the custom date range year by year, and adding up the partial results at the end (year-by-year).

For this purpose, first we set the custom range from 1700 to 2013. We have used this time span because data before 1700 are practically inexistent (only 49 records were found in the 1000-1700 range). This data collection process was carried out in May 2014.

Secondly, we directly queried the English version of Google Scholar by means of an empty query search, (that is, leaving the search box empty) filtering by single years (year-by-year) from 1700 to 2013, and gathering the estimated number of results, also called Hit Count Estimates (HCE).[2] After this, the partial results obtained for each year were added together.

Articles and case laws need to be measured by means of independent queries since these document types are not combined in the SERP. Additionally, to test the potential influence of citations and patents in the size of Google Scholar, we retrieved the following data for each year:
- All documents (records + citations + patents);
- Records + citations;
- Records + patents;
- Only records, excluding citations and patents.

All queries in Google Scholar were manually entered via http, and this task was equally distributed among all the authors of this article.

Finally, direct queries were performed on Microsoft Academic Search as well as on Web of Science Core Collection and Scopus, with the aim of gathering data about the current size of these databases, so as to compare them to those obtained previously for Google Scholar:
- Microsoft Academic Search: a direct query was performed via the "year" command on <academic.research.microsoft.com>.
- Web of Science Core Collection: the size was obtained from the basic search interface, specifying the year in the "Year published" field. Data concerning language and type of document were gathered as well.
- Scopus: the size was obtained from the advance search interface, specifying the year in the "Pubyear is" field.

For these three databases, a query per year (from 1700 to 2013) and a global query from 1700 were performed (data were collected in May 2014).

**b) Absurd query**

The last method proposed is based on some of the features of Boolean logic that are supported in Google Scholar's search box. In this case, the goal is to compose a query that somehow requires Google Scholar to return all its records. Although nowhere in the official documentation is it stated that such a query exists, we ran test queries using the following syntax: <common_term -site:non-existent_site>

The idea behind this is to query the number of occurrences of a very common term (likely to appear in almost all written records), and to filter out its appearances in a non-existent web site, which means that we are implicitly selecting every existing site. For example: <a -site:ssstfsffsdfasdfsf.com>, or <1 -site:ssstfsffsdfasdfsf.com>. The reason for including a term before the "-site" command is that this command does not work on its own.

As in the case of the direct query method, the queries were performed in two different ways: a) setting the custom range from 1700 to 2013, and b) running a query for each year and adding the annual results together at the end. All these queries were run on Google Scholar both for Article and Case laws (including and excluding citations and patents) in June 2014.

**4. Results**

Various estimates of the size of Google Scholar, as calculated from each of the three procedures outlined above, are offered below.

### 4.1. Estimates from empirical data

In Table 2 we can observe the median, and the number of studies that make up the empirical set for each unit of study: number of documents (number of original source documents) and unique citing documents (those created from unique cited references). The complete results obtained from the empirical data are available in Appendix I.

**Table 2. Correction factor from empirical studies on Google Scholar coverage**

| UNIT OF STUDY | MEDIAN | N* |
|---|---|---|
| Number of documents | 3 | 8 |
| Unique citing documents | 2.4 | 9 |

* The studies with fewer than 10 documents in the sample of WoS have not finally been considered since they are not representative enough.

We might assume (based on the empirical studies referenced in Appendix I as well as the data provided in Table 2) that, on a general level, GS could triple the contents of WoS (that is, a round correction factor of 3), regardless of the different English content distribution, and the different document type distribution of each of these databases. This is particularly significant as it implies that the size comparison is not influenced by the biases in WoS towards the English language (bias should not be regarded as a negative concept here. It is used to reflect the elevated percentage of documents in such language in WoS as compared to the percentage of documents in other languages, regardless of quality or impact) and the article document type.[3]

We have ascertained from the empirical studies that the proportion of English documents in GS is around 65% (Appendix III). This would mean that, for documents in English, GS does not triple the number of WoS documents, but it probably does for documents in other languages. Nonetheless, knowing the general size correction factor (equal to 3), we do not need to worry about language distribution in GS for calculating estimates.

Therefore, by inference from samples of previously conducted empirical studies, we conclude that we may simply multiply the size of WoS by three. As WoS currently has about 57 million records (as of May 2014), we may estimate around 171 million records for GS. These data and relationships are expressed formally in Table 3.

**Table 3. Size relationships between Web of Science and Google Scholar**

| EQUATION | OBSERVATIONS |
|---|---|
| **3 * WoS = GS [1]** | We apply a correction factor of 3 (GS triples WoS) |
| **WoS = WoSe*0.9 + WoSo*0.1 [2]** | WoSe: English content in WoS<br>WoSo: WoS content in other languages |
| **GS = GSe*0.65 + GSo*0.35 [3]** | GSe: English content in GS (65% from empirical data)<br>GSo: GS content in other languages |
| **3 * (WoSe*0.9+WoSo*0.1) = GSe*0.65 + GSo*0.35** | Substituting [2] and [3] in [1] |
| **WoS = 57 million documents;<br>WoSe = 51.3 million documents;<br>GS = 171 million documents;** | WoS currently gives approximately 57 million records |

Source: prepared by the authors. Data as of May 2014

If we consider the 65% of documents in GS that are in English, estimated from the empirical data, we obtain about 111.15 million documents in English (as of May 2014), slightly higher than the 99.3 million estimated by Khabsa & Giles (as of January 2013).

### 4.2. Estimates from direct empty query

The second strategy proposed in this study consists of querying the databases directly through their search interface, using both a wide year range query and year-by-year queries.

*Wide year range query*

First, we performed a query on Google Scholar, selecting the range 1700 to 2013 (via the custom year range option). Unfortunately, this procedure failed, returning only 596,000 documents. We performed some other partial queries (considering specific periods) in order to corroborate this error (Table 4).

**Table 4. Custom range option error in Google Scholar**

| PERIOD | HCE |
|---|---|
| 1700-2013 | 596,000 |
| 1750-2013 | 567,000 |
| 1800-2013 | 552,000 |
| 1850-2013 | 566,000 |
| 1900-2013 | 541,000 |
| 1950-2013 | 617,000 |
| 2000-2013 | 693,000 |

The results displayed in Table 4 not only show a low number of results for such wide timeframes, but also highlight serious inconsistencies. For example, in the time span "2000-2013", the system retrieves more documents than in longer periods.

However, if we execute the query introducing only 1 year in the custom range, the results seem to be more plausible. For example, for the year "1900", we obtain 141,000 results and for the year "2000", we get 2,410,000 results. Therefore, in order to solve this problem, a year-by-year analysis is required.

*Year-by-year queries*

The article search from 1700 to 2013 returns 99.8 million results in Google Scholar (64.87 million documents written in English if we apply the 65% rule of thumb discussed above). The comparative data from the databases (WoS, Scopus, MAS and GS) is shown in Figure 1.

**Figure 1. Number of documents in GS, MAS, Scopus and WoS (1700-May 2014)**

The evolution of these four databases (from 1800 to 2013) is shown in Figure 2.

**Figure 2. Google Scholar, Microsoft Academic Search, Web of Science, and Scopus (1800–2013)**

Figure 2 emphasises the predominance of Google Scholar during practically the entire period (over 200 years), except in the 1970s, where its performance is similar to WoS.

This prevalence appears to accelerate again in the last decade of the twentieth century and the first years of the twenty-first, except for some inconsistencies identified in 2009-2010 (from 3,110,000 to 1,840,000 records) and 2011-2012 (from 3,230,000 to 2,410,000), probably due to internal changes within the search engine.

In any case, it should be noted that the total result of 99.8 million documents in Google Scholar includes both patents and citations. If we exclude these two types of documents from the query, the results fall dramatically to 80.5 million. In Figure 3, we present the totals disaggregated by records (80.69%), citations (18.38%) and patents (0.92%), since 1700.

**Figure 3. Composition of Google Scholar results: records, citations and patents**

However, these results should be taken with extreme caution, because the hit count estimates offered by Google Scholar for these queries are far from accurate. Considering the 314 years for which calculations were performed (from 1700 to 2013), we found inconsistencies between the complete query (records + citations + patents) and a partial query (records + citations) in 122 years, finding more results in the latter than in the former. In short: excluding patents, the system sometimes (39% of the time) retrieves more results for the partial query than with the complete query.

In Table 5 we show some examples of years where these inconsistencies occur. Due to the order of magnitude of the recent data (millions of results since the end of the 20$^{th}$ century), error rates reach unsustainable values.

**Table 5. Inconsistencies in Google Scholar queries for patents and citations**

| YEAR | QUERY | | DIFFERENCE |
| --- | --- | --- | --- |
| | RECORDS + CITATIONS + PATENTS | RECORDS + CITATIONS | |
| 2013 | 4,070,000 | 4,150,000 | -80,000 |
| 2010 | 1,840,000 | 2,020,000 | -180,000 |
| 2009 | 3,110,000 | 3,230,000 | -120,000 |
| 2007 | 2,990,000 | 3,110,000 | -120,000 |
| 2006 | 3,000,000 | 3,050,000 | -50,000 |
| 2005 | 2,920,000 | 2,950,000 | -30,000 |
| 2004 | 2,860,000 | 2,930,000 | -70,000 |
| 2002 | 2,620,000 | 2,720,000 | -100,000 |
| 2000 | 2,410,000 | 2,550,000 | -140,000 |

As regards citations, the total figure obtained previously (18,355,380 citations) is higher than expected. The accuracy is greater than for patents, producing only 8 errors in 314 years, and focused in a narrow time span of 20 years: 1969 (59,000 more records obtained through the partial query), 1970 (36,000), 1971 (52,000), 1975 (19,000), 1976 (11,000), 1978 (56,000), 1982 (10,000) and 1988 (40,000).

Google Scholar includes one more document type apart from the "Articles" category (composed of records, citations, and patents): this is case law from the Supreme Court of the United States of America, which also includes citations.

A year-by-year analysis of the number of case law results from 1700 was performed, in a similar way as for articles, both including and excluding citations in the search results. In Figure 4 we can observe the evolution since 1800 (from 1700 to 1799 Google Scholar retrieves only 408 documents).

**Figure 4. Number of case law results per year (1800-2013)**

A total of 26,510,689 case law results (31.3%) and citations to case laws (68.7%) were obtained since 1700. We should highlight the great differences between the number of case law results, and the number of citations to case laws, during the second half of the nineteenth century and the first half of the twentieth. After that, and until 2013, both types share very similar behaviour.

If case law results and their citations are included in the calculation of Google Scholar's total size, the global figure rises to 126,341,609 documents, a figure about twice the size of WoS (56.9 million).

**4.3. Estimates from direct absurd query**

This method is applied under three different approaches: without a temporal filter, using the custom range (from 1700 to 2013), and finally by means of a year-by-year analysis. Particular screenshots and tests are available in the supplementary material (Appendix V).

*Without temporal filter*

The query <a -site:ssstfsffsdfasdfsf.com> was tested, obtaining 102,000,000 documents (partial query excluding patents and citations) and 154,000,000 documents (excluding only patents). An attempt to include both patents and citations (theoretically the query with a higher count) resulted in an error message informing us that there had been technical problems in delivering results (the reason behind may be related to the time needed to resolve the query). Therefore, this query was discarded.

After this, the query <1 -site:ssstfsffsdfasdfsf.com> was applied to Google Scholar's articles category, excluding patents and citations (obtaining 127,000,000 results), excluding only patents (158,000,000), and including all documents (170,000,000). Moreover, this same query, applied to case laws (with citations), returned 4,550,000 results, far from the 26.5 million obtained through the year-by-year analysis discussed in the previous method.

*Wide year range (1700 to 2013)*

In this case, the query <1 -site:ssstfsffsdfasdfsf.com> retrieved 176 million results for articles and 4.3 million for case laws. As regards the query <a -site:ssstfsffsdfasdfsf.com>, this time it worked, returning 160 million articles and 6.8 million case law results.

*Year-by-year (1700 to 2013)*

Finally, the absurd query <1 -site:ssstfsffsdfasdfsf.com> was performed each year from 1700 to 2013.

The final count gives an overall figure of 169.5 million articles and 3.4 million case law results, far higher than the figures obtained from the year-by-year direct empty query (99.8 million articles and 26.5 case law results, respectively).

Finally, it should be mentioned that, although the results obtained are different from those achieved previously with the year-by-year direct empty query, these results correlate. Pearson correlation (r) between the number of articles per year in both methods (empty and absurd query) is r = .93 (for case laws it is r = .71).

This confirms that Hit Count Estimates from Google Scholar are not useful for achieving accurate performance for individual queries, but are useful for making performance comparisons.

### 4.4. Results overview

Finally, Table 6 summarises the results obtained for each method.

**Table 6. Summary of Google Scholar size estimates**

| METHOD | GS SIZE ESTIMATE | COMMENTS |
|---|---|---|
| **A. Data from empirical studies** | 171 million | This method assumes GS triples the size of WoS |
| **B.1. Empty query (wide year range)** | 1.2 million | This method applies an empty query in setting the custom range from 1700 to 2013. |
| **B.2. Empty query (year-by-year)** | 126.3 million | This method applies an empty query in a year-by-year analysis from 1700 to 2013 including articles (99.8 million) and case law (26.5 million). |
| **C.1. Absurd query (total)** | 174.5 million | 170 million articles and 4.5 million case law results |
| **C.2. Absurd query (wide year range)** | 180.3 million | This method applies an absurd query setting the custom range from 1700 to 2013, both for articles (176 million) and case law (4.34 million) |
| **C.3. Absurd query (year-by-year)** | 172.9 million | This method applies an absurd query in a year-by-year analysis from 1700 to 2013, obtaining 169.5 million articles and 3.4 million case law results. |

### 5. Discussion

In this section, we discuss the advantages and disadvantages of each method as well as the disparity of the results, as observed in Table 6.

*Method A: data from empirical studies*

The estimates from method A (data from empirical studies) have some considerable shortcomings since it is difficult to synthesise empirical results from studies managing different sample sizes, obtained in different periods, and even using different methods of sample selection. Furthermore, sample studies may not be representative since the citation network may not be uniformly distributed.

Moreover, the disciplines or specialties under study are varied. We must remember that the communication patterns and dynamics of scientific publications are very different from one discipline to another, and this can seriously affect the results. Additionally, in some of these studies, databases are searched by author names and authority control may affect the results (e.g., searching by last name + initial).

Finally, this method may be affected by the age of documents analysed in the sample of empirical studies since there are huge historic trends in the databases coverage, i.e. the correction factor may be higher or lower depending on the year of publication of the documents. In this sense, we should emphasise that these calculations are based on samples and averages and it is supposed to be an estimation; the more articles included in the correction factor calculation the better to minimize this bias.

Conversely, this method has the advantage of not being affected by estimates taken from biased databases (towards the English language and the article document type), as is the case of the method followed by Khabsa and Giles (2014), which depends on the coverage of Microsoft Academic Search. This is the reason why these authors intended to measure only the number of English scholarly documents (journal articles), which constitute only a portion of Google Scholar.

Focusing exclusively on the size of the English scholarly world, Khabsa and Giles found 99.3 million documents (as of January 2013) whereas with the empirical method we found 111.15 million documents (as of May 2014). Considering the time difference between both studies (16 months), and the growth rate of Google Scholar, the difference is unexpectedly small, despite some inconsistencies found in Khabsa and Giles' method (especially due to the analogy used in employing the Lincoln-Petersen estimation method).

Nonetheless, Khabsa and Giles' method is novel and promising, since it is an external procedure based on the collection of documents that cite a sample of articles, which has several advantages (such as the fact that the search engine is forced to query its entire database to find all documents that match a citation to any of the documents of the sample).

*Method B: empty query*

The methods based on the use of a direct query (empty or absurd, wide year range or year-by-year) on the academic search engine raise a number of unavoidable methodological issues:
- Validity: the extent to which the search engine returns, from our query, what we really want to measure, i.e., the number of unique records indexed in Google Scholar.
- Precision: the extent to which the search engine returns data for our query that corresponds with the reality of its catalogued universe. We rely on hit count estimates (Google explicitly states "about xxx results"), which are affected by unknown rounding routines.
- Reliability: the extent to which the search engine returns, from our query, similar results under consistent conditions.

The fact that there is no API for Google Scholar, and that it only displays the first 1,000 results, prevents us from performing large-scale empirical studies of these issues. This issue affects the replicability of this kind of studies as well. In this case, although the procedure may easily be repeated, data may vary because retrospective indexing can occur in Google Scholar; this is an issue that should be taken into account.

Furthermore, the influence of mistakes in the bibliographic description of records should be considered as well. In 2005-2007, according to Jacsó (2008), there were major mistakes (mainly related to documents with wrong dates of publication and authorship, and duplicates due to not having correctly linked different versions).

Today, these errors have been solved for the most part, although some persist. Among them, we should highlight an unknown percentage of publications with no date. Since the direct method depends on the specification of a publication date, the results obtained under this method are an underestimation, and may explain the lower results of method B (both B.1 and B.2).

Otherwise, we should acknowledge the procedure used to obtain the number of hits (Hit Count Estimates) is inaccurate (Jacsó 2008; Uyar, 2009). On the one hand, the search command "site" is not exhaustive; on the other hand, the number of hits recalled is a rounded value.

Citations present an additional problem since it has been confirmed that not all retrieved citations exactly match their own definition (that is, records that Google Scholar has not been able to find on the web, and for which only a bibliographical description is provided). In some cases, the same article appears as a record and as a citation because the system has been unable to detect the fact that they are different versions of the same document, and therefore they have not been combined. The presence of duplicates directly affects the global Hit Count Estimates.

Considering the limitations previously exposed (which confirm that number of hits does not match the number of documents, being thus only an approximation), in the case of empty query using a wide year range (method B.1) this method must be dismissed as inaccurate. The results provided in Table 4 confirm a dysfunction of custom year option.

In the case of the empty query using year-by-year query (method B.2), 99.8 million items (May 2014) were obtained. These results are similar to those obtained by Ortega (2014), who, using the same method, obtained 94.73 million articles (as of December 2013). Given the growth rate of Google Scholar, the difference of 5 million records can be considered to be within normal limits. By contrast, in the case of case law, the difference is greater because Ortega obtained 14.57 million, far below the 26.5 million case law results obtained in this study.

Moreover, the highly significant problems identified in recent years, precisely when world output actually accelerates, account for the dangerous instability of Google Scholar (Aguillo 2011; Orduna-Malea and Delgado López-Cózar 2014; Martín et al. 2014 ) and search engine hit count estimates in general (Jacsó 2006).

The fact that the number of records decreases from year to year obviously does not mean a lower production in those years, but that the academic search engine has made internal adjustments, deleting duplicates, fixing bugs, among other technical issues. These data highlight the similar sizes of WoS, Scopus and MAS from 2000 onwards, until 2010, when the coverage of Microsoft Academic Search starts to falter (Orduna-Malea et al. 2014).

*Method C: absurd query*

As regards the absurd query (method C), though it is closer in nature to method B (an internal query of the database), it produced significantly different results (169.5 million articles), closer to those obtained by method A. It is possible that the dysfunction between the selection of Citations and Patents and the HCE obtained (see Table 6) may influence each method differently.

In this sense, a limitation in the detection of case laws has been identified. The year-by-year direct absurd query only retrieves 3.4 million case law results.

In order to ascertain the reason for the differences in the raw data between these two methods (both based on year-by-year querying the Google Scholar database), the search results that the absurd query generated were analysed in more detail. Thus, we have identified the following weaknesses:

a) The absurd query does not retrieve citations (although this option is checked in the search options), both in the case of articles and case laws (see Figure 4 for the proportion of citations to case law). Conversely, the empty query retrieves citations, as was noted previously. This may explain the differences between these queries in the year-by-year results for case law.
b) The Hit Count Estimates (HCE) present serious inconsistencies in the activation/deactivation of the citation inclusion feature. For example, filtering by the year 1840, and with the option "Citations" deactivated, we obtained 39 results. At the same time, activating the "Citations" option we got only 9 results. Moreover, the system only retrieved 2 records, even though the HCE said there were 7 (see example in Appendix V).
c) We discovered the existence of empty and false SERPs (Search Engine Results Pages). For example, applying the absurd query for a given year, we got a HCE of 132, but accessing the 6th SERP we verified that it was empty. Setting the system to retrieve up to 20 results per SERP (the maximum allowed in GS), the 6th SERP should show results from 101st to 120th, and never an empty page. Moreover, we observed that a 15th SERP was created (unnecessary with 132 results). When we clicked on this SERP, the system not only still displayed an empty SERP but the HCE increased as well (from 132 to 521) (see example in Appendix V). These shortcomings are, as of yet, unexplained. It is possible that the system goes into a loop when trying to answer a query of this type.

Yet it is surprising that the final figures seem logical and coherent, and close to those achieved by other methods (unlike what happens with the empty query method). This is probably because the search engine is forced to check the entire database to answer the query, as the time responses suggest.

**6. Conclusions**

Considering only the results obtained within the period 1700-2013, the size of Google Scholar may be estimated to be around 170-175 million unique records (regardless the number of versions of each record). From our experience, we conjecture that there is currently a maximum of 10% of internal errors (excluding undated documents) that would leave the final figure at about 160-165 million, although we must test this hypothesis empirically in future research.

However, all methods show great inconsistencies, limitations and uncertainties. The external method (estimates based on the comparative differences between databases), has the advantage of not being affected by comparisons with other biased databases, although the estimate is very rough and imprecise as it is synthesised from diverse empirical results. The internal method based on the empty query seemed to be more plausible a priori (being based on directly querying Google Scholar), but the results are unexpectedly low (showing inaccuracies in the custom range function and effects of the existence of undated publications). The internal method based on an absurd query does not retrieve citations and creates empty SERPs, thus affecting the hit count estimates.

Surprisingly, even though all methods seem invalid for various and diverse reasons, the external method and internal method based on absurd query (with all variants considered) return similar results despite being of a different nature, reinforcing the validity of the estimation performed.

Logically this matter should be resolved by simply asking Google Inc. for this information. Their answer would render all our efforts and resources dedicated to finding this elusive "golden fleece" unnecessary. Nonetheless, the lack of information from Google encourages speculation and forces researchers to make conjectures about the real size of Google Scholar by designing and testing new quantitative methods. This in turn will allow a better understanding of Google Scholar.

## 7. Notes

1. http://thewebindex.org
2. The Custom range option appears after a query is submitted in the search box of Google Scholar. The user can also access to the advanced search option to set the year range. Moreover, we can execute this query directly on the browser via http as well. Once we obtain the first results via hit count estimates, we can generate new queries without introducing any keyword in the search box, and only selecting the time span required. This is the procedure followed in this study.
3. Additional information about the biases of WoS towards English and article document type is available in the supplementary material (Appendix V).

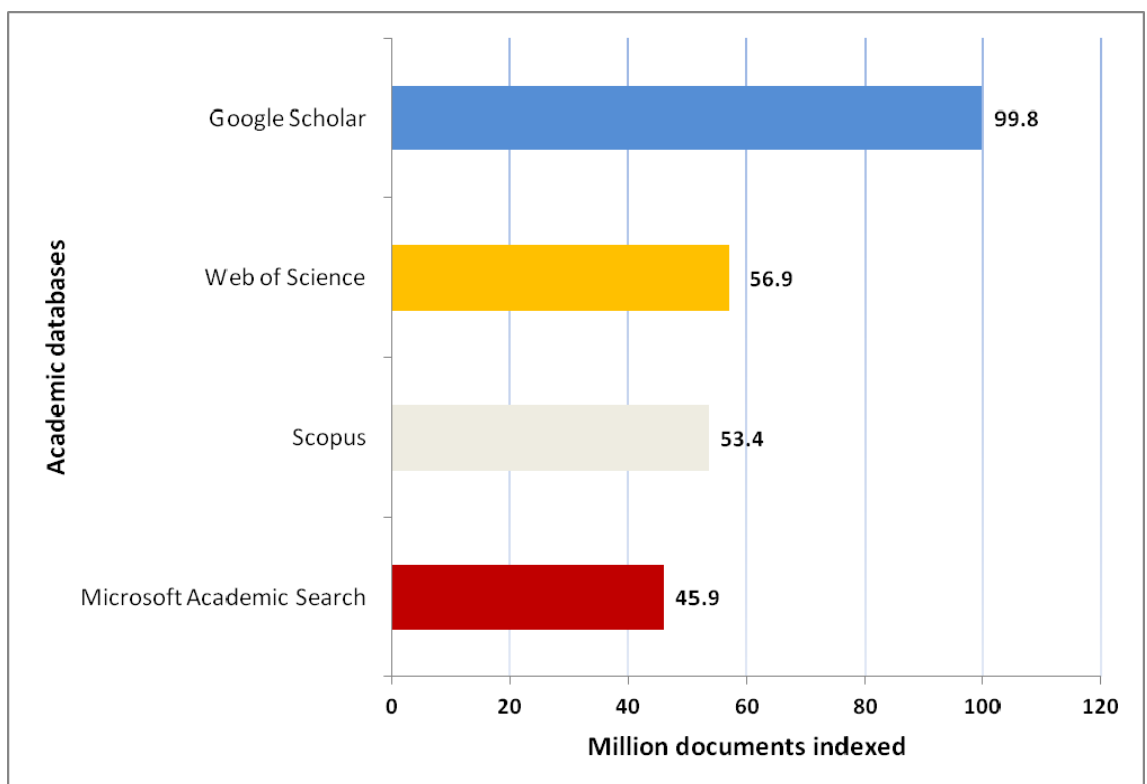

**Figure 1. Number of documents in GS, MAS, Scopus and WoS (1700-May 2014)**

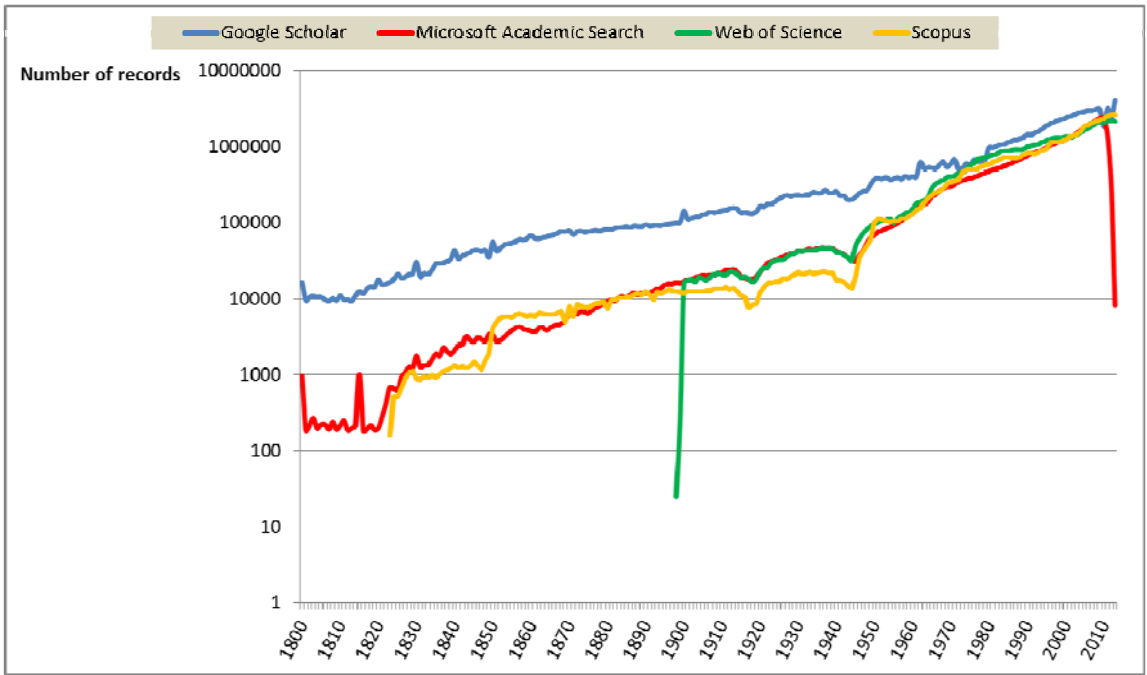

**Figure 2. Google Scholar, Microsoft Academic Search, Web of Science, and Scopus (1800–2013)**

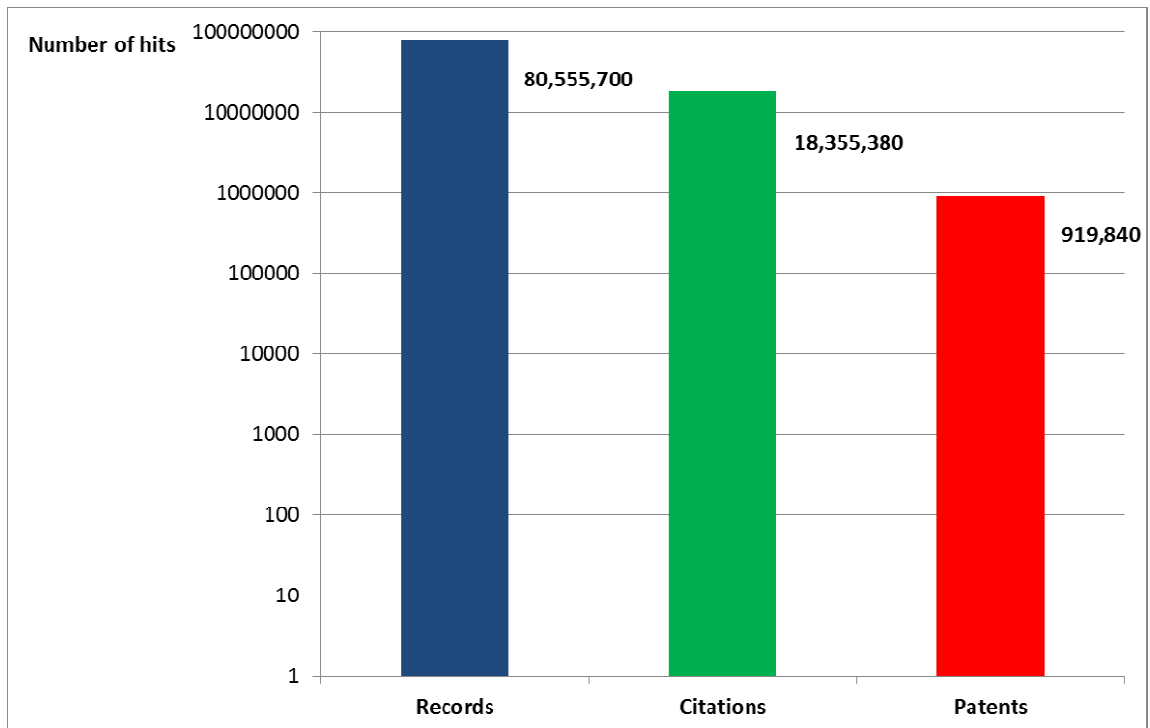

**Figure 3. Composition of Google Scholar results: records, citations and patents**

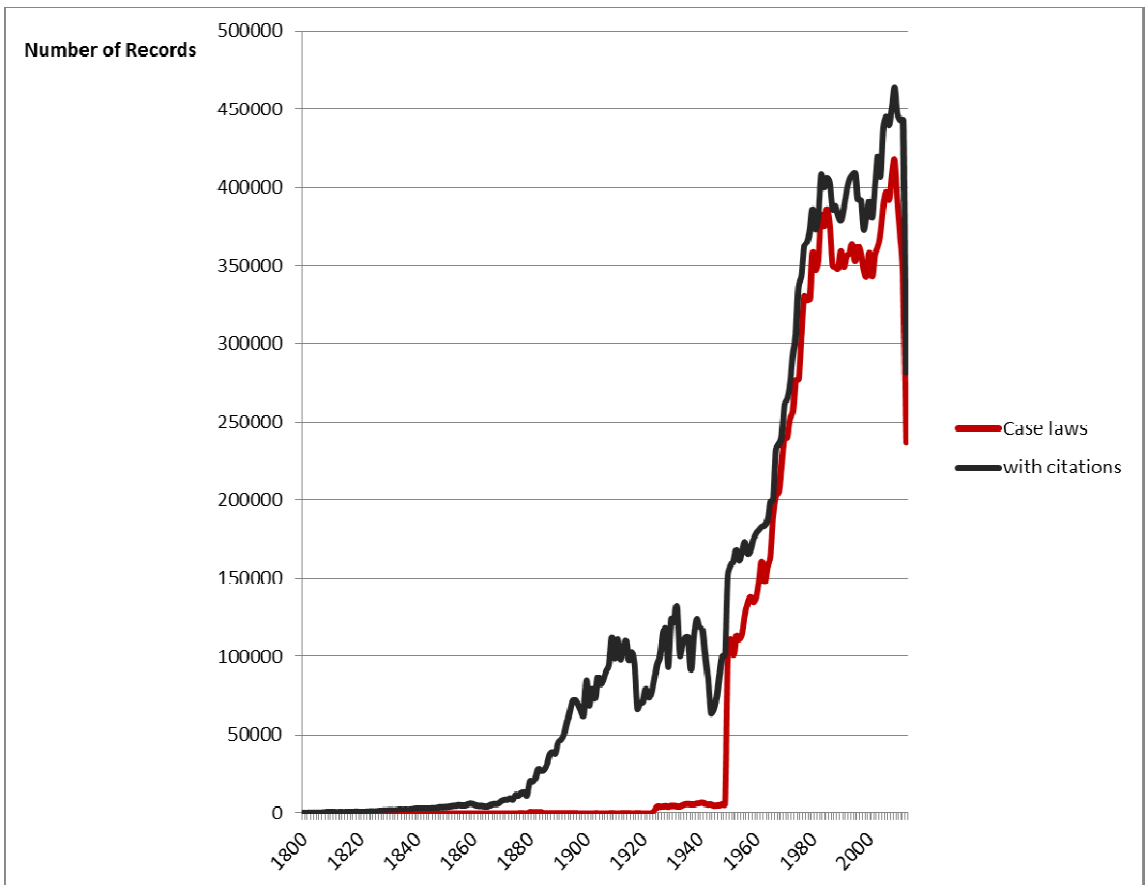
**Figure 4. Number of case law results per year (1800-2013)**